\documentclass[conference, 10pt]{IEEEtran}
\IEEEoverridecommandlockouts
\usepackage{amsmath, mathrsfs,amsfonts}
\usepackage{amsfonts}
\usepackage{amssymb}
\usepackage{graphicx, amsthm}
\usepackage[letterpaper]{geometry}
\usepackage{hyperref}

\usepackage{epsfig}
\usepackage{graphicx}
\graphicspath{{./figures/}}
\usepackage{amsmath}
\usepackage{times}
\usepackage{amssymb}
\usepackage{amsthm}
\usepackage{amsfonts}
\usepackage{algorithmic}

\newtheorem{definition}{Definition}

\newcommand{\F}{\mathbb{F}_q}

\newcommand{\beqn}{\begin{equation}}
\newcommand{\eeqn}{\end{equation}}

\newcommand{\ie}{{\it i.e.}}

\begin{document}
\title{Interfacing network coding with TCP: an implementation}
\author{
\IEEEauthorblockN{Jay~Kumar~Sundararajan\authorrefmark{1}, Szymon~Jakubczak\authorrefmark{1}, Muriel~M\'edard\authorrefmark{1}, Michael Mitzenmacher\authorrefmark{2}, Jo\~ao Barros\authorrefmark{3}}
\IEEEauthorblockA{
\begin{tabular*}{\textwidth}{p{2.2in}p{2.1in}p{2.3in}}
\ &\ &\ \\
{\centering \authorrefmark{1}Dept. of EECS\\ Massachusetts Institute of Technology,\\ Cambridge, MA 02139, USA\\ \{jaykumar, szym, medard\}@mit.edu\\} & 
{\centering \authorrefmark{2}School of Eng. and Appl. Sciences\\Harvard University,\\Cambridge, MA 02138, USA\\
michaelm@eecs.harvard.edu\\} & 
{\centering \authorrefmark{3}Instituto de Telecomunica\c{c}\~oes\\
Dept. de Engenharia Electrot\'ecnica e de Computadores\\
Faculdade de Engenharia da Universidade do Porto, Portugal\\
jbarros@fe.up.pt\\}
\end{tabular*}
}
}
\maketitle

\begin{abstract}
In previous work (`Network coding meets TCP') we proposed a new protocol that interfaces network coding with TCP by means of a coding layer between TCP and IP. Unlike the usual batch-based coding schemes, the protocol uses a TCP-compatible sliding window code in combination with new rules for acknowledging bytes to TCP that take into account the network coding operations in the lower layer. The protocol was presented in a theoretical framework and considered only in conjunction with TCP Vegas. In this paper we present a real-world implementation of this protocol that addresses several important practical aspects of incorporating network coding and decoding with TCP's window management mechanism.  Further, we work with the more widespread and practical TCP Reno.  Our implementation significantly advances the goal of designing a deployable, general, TCP-compatible protocol that provides the benefits of network coding.
\end{abstract}

\section{Introduction}
The Transmission Control Protocol (TCP) was originally developed for wired networks. Since wired networks have very little packet loss on the links and the predominant source of loss is buffer overflow due to congestion, TCP's approach of inferring congestion from losses works well. In contrast, wireless networks are characterized by packet loss on the link and intermittent connectivity due to fading. TCP wrongly assumes the cause of these link losses to be congestion, and reduces its transmission rate unnecessarily, leading to low throughput. These problems of TCP in wireless networks are very well studied, and several solutions have been proposed (see \cite{rangwala} and references therein for a survey).

In past work we proposed a new protocol called TCP/NC \cite{infocom09} that incorporates network coding inside the TCP/IP protocol stack with the aim of improving TCP throughput in wireless networks. The interface of TCP with network coding can be viewed as a generalization of previous work combining TCP with Forward Erasure Correction (FEC) schemes \cite{Brockners99}. As opposed to coding only at the source, the protocol of \cite{infocom09} also allows intermediate nodes in the network to perform re-encoding of data. It is thus more general than end-to-end erasure correction over a single path, and can therefore, in principle, be used in multipath and multicast scenarios to obtain throughput benefits.

In the current work, we present a real-life network coding implementation based on the mechanism proposed in \cite{infocom09}. The main contributions of this paper are as follows:
\begin{enumerate}
	\item We explain how to address the practical problems that arise in making the network coding and decoding operations compatible with TCP's window management system, such as variable packet length, buffer management, and network coding overhead.
	\item We demonstrate the compatibility of our protocol with the widely used TCP Reno;  the original proposal of \cite{infocom09} considered only TCP Vegas.
	\item We present experimental results on the throughput benefits of the new protocol for a TCP connection over a single-hop wireless link. Although currently our experiments only study behavior over a single hop, this restriction is not mandatory and the evaluation of the protocol over arbitrary topologies will be addressed elsewhere.
\end{enumerate}

Before beginning, we explain the implications of this new protocol for improving throughput in wireless networks.  There has been a growing interest in approaches that make active use of the intrinsic broadcast nature of the wireless medium. In the technique known as opportunistic routing \cite{exor}, a node broadcasts its packet, and if one of its neighbors receives the packet, that node will forward the packet downstream, thereby obtaining a diversity benefit. If more than one of the neighbors receive the packet, they will have to coordinate and decide who will forward the packet.

The MORE protocol \cite{more} proposed the use of intra-flow network coding in combination with opportunistic routing. The random linear mixing (coding) of incoming packets at a node before forwarding them downstream was shown to reduce the coordination overhead associated with opportunistic routing. Another advantage is that the coding operation can be easily tuned to add redundancy to the packet stream to combat erasures. Such schemes can potentially achieve
capacity for a multicast connection \cite{RLC}.

Typical implementations use batches of packets instead of sliding windows, and are generally therefore not compatible with TCP. ExOR uses batching to reduce the coordination overhead, but as mentioned in \cite{exor}, this interacts badly with TCP's window mechanism. MORE uses batching to perform the coding operation. In this case, the receiver cannot acknowledge the packets until an entire batch has arrived and has been successfully decoded. Since TCP performance heavily relies on the timely return of ACKs, such a delay in the ACKs would affect the round-trip time calculation and thereby reduce the throughput.

Opportunistic routing also leads to reordering of packets, which is known to interact badly with TCP, as reordering can cause duplicate ACKs, and TCP interprets duplicate ACKs as a sign of congestion. The work of \cite{infocom09} addresses both these issues. It proposes a TCP-compatible sliding window coding scheme in combination with a new acknowledgment mechanism for running TCP over a network coded system. The sender would transmit a random linear combination of packets in the TCP congestion window. The new type of ACK allows the receiver to acknowledge every linear combination (degree of freedom) that is linearly independent from the previously received linear combinations. The receiver does not have to wait to decode a packet, but can send a TCP ACK for every degree of freedom received, thus eliminating the problems of using batchwise ACKs.

It was shown in \cite{infocom09} that if the linear combination happens over a large enough finite field, then every incoming random linear combination will, with high probability, generate a TCP ACK for the very next unacknowledged packet in order. This is because the random combinations do not have any inherent ordering. The argument holds true even when multiple paths deliver the random linear combinations. Hence the use of random linear coding with the acknowledgment of degrees of freedom can potentially \emph{address the TCP reordering problem for multipath opportunistic routing schemes}.  By presenting an implementation of the TCP/NC protocol of \cite{infocom09}, this work provides a  way of combining TCP with network-coding-based multipath opportunistic routing protocols such as MORE.

The rest of the paper is organized as follows. Section \ref{sec:overview} summarizes the protocol proposed in \cite{infocom09} and provides an overview of the system modifications required. Sections \ref{sec:sender} and \ref{sec:receiver} describe the sender side and receiver side modules in detail. In section \ref{sec:factors}, we discuss the parameters defined in the algorithm and how they affect the performance. Section \ref{sec:results} presents the results obtained from the experiment. Finally, conclusions and possible future work are presented in Section \ref{sec:conc}. 

\section{An overview of the protocol}\label{sec:overview} 
\subsection{The architecture}
\begin{figure}%
\centering
\includegraphics[width=\columnwidth]{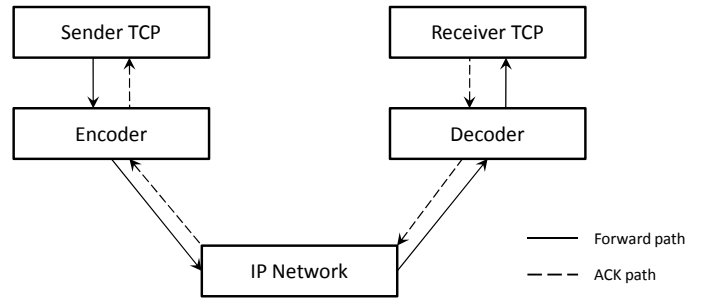}%
\caption{Overview of the new protocol}%
\label{fig:blockdiag}%
\end{figure}

The TCP/NC protocol introduces a network coding layer between TCP and IP in the protocol stack as shown in Fig. \ref{fig:blockdiag}, where an encoder module lies on the sender side and a decoder module lies on the receiver side. 
Although it is not shown in this figure, the system can be generalized to have re-encoding inside the network, in a manner similar to MORE \cite{more}, but above the IP layer. 

\subsection{The operations}
The encoder buffers packets generated by TCP and for every arrival from TCP, it transmits $R$ random linear combinations of the buffered packets on average, where $R$ is the redundancy factor.  The contents of a coded packet represent a linear combination of the original uncoded packets; to convey the combination requires an additional network coding header that is added to the coded packet. The original uncoded packets are retained in the encoding buffer until an appropriate TCP ACK arrives from the receiver side. 

The purpose of adding redundancy is to separate the loss recovery aspect from the congestion control aspect. Losses can now be recovered without forcing TCP retransmissions and the associated congestion window size reductions. The amount of redundancy that is added depends on how lossy the network is. For instance, a 10\% loss rate means that the rate at which equations are sent into the network should be roughly 10\% more than the rate at which packets enter the encoder from TCP. This would ensure that the number of equations reaching the receiver will match the number of packets entering the encoder.

On the decoder side, upon receiving a new linear combination the decoder places it in a decoding buffer, appends the corresponding coefficient vector to the decoding matrix, and performs Gaussian elimination. This process helps identify the ``newly seen'' packet (if it exists). The notion of a node seeing a packet was defined in \cite{infocom09} and is repeated here for the reader's convenience. Packets are treated as vectors over a finite field $\F$ of size $q$. The $k^{th}$ packet that the source generates is said to have an \emph{index} $k$ and is denoted as $\mathbf{p_k}$.

\begin{definition}[Seeing a packet]\label{def:seen}
  A node is said to have \emph{seen} a packet $\mathbf{p_k}$ if it has enough information to compute a linear combination of the form $(\mathbf{p_k} + \mathbf{q})$, where $\mathbf{q} = \sum_{\ell > k} \alpha_\ell \mathbf{p_\ell}$, with $\alpha_\ell \in \F$ for all $\ell > k$. Thus, $\mathbf{q}$ is a linear combination involving packets with indices larger than $k$. 
\end{definition}

Alternately, we can say that a packet is seen if after Gaussian elimination of the coefficient matrix, the packet corresponds to one of the pivot columns (a column that contains the first non-zero element in some row). The decoder then sends a TCP ACK to the sender requesting the first unseen packet in order. Thus, the ACK is a cumulative ACK like in conventional TCP. The Gaussian elimination may result in a new packet being decoded. In this case, the decoder delivers this packet to the receiver TCP. Any ACKs generated by the receiver TCP are suppressed and not sent to the sender. These ACKs may be used for managing the decoding buffer.

\subsection{Clean interface with TCP}
An important point to note is that the introduction of the new network coding layer does not require any change in the basic features of TCP. As described above, the network coding layer accepts TCP packets from the sender TCP and in return delivers regular TCP ACKs back to the sender TCP.  On the receiver side, the decoder delivers regular TCP packets to the receiver TCP and accepts regular TCP ACKs. Therefore, neither the TCP sender nor the TCP receiver sees any difference looking downwards in the protocol stack\footnote{Certain advanced features of TCP may require some changes. See Section \ref{sec:misc} for further discussion.}. The main change introduced by the protocol is that the TCP packets from the sender are transformed by the encoder by the network coding process. This transformation is removed by the decoder, making it invisible to the TCP receiver. On the return path, the TCP receiver's ACKs are suppressed, and instead the decoder generates regular TCP ACKs that are delivered to the sender. Again, the sender does not need to be aware of the coding layer below. This interface allows the possibility that regular TCP sender and receiver end hosts can communicate through a wireless network even if they are located beyond the wireless hosts.

We now discuss some of the practical issues that arise in designing an implementation of the TCP/NC protocol compatible with real TCP/IP stacks. These issues were previously not considered in the idealized setting of \cite{infocom09}. We show that it is possible to implement a TCP-aware network-coding layer that has the property of a clean interface with TCP, as described above.

\section{Sender side module}\label{sec:sender}
\subsection{Forming the coding buffer}
The description of the protocol in \cite{infocom09} assumes a fixed packet length, which allows all coding and decoding operations to be performed symbol-wise on the whole packet. That is, in \cite{infocom09} an entire packet serves as the basic unit of data (\ie, as a single unknown), with the implicit understanding that the exact same operation is being performed on every symbol within the packet. The main advantage of this view is that the decoding matrix operations (\ie, Gaussian elimination) can be performed at the granularity of packets instead of individual symbols. Also, the ACKs are then able to be represented in terms of packet numbers. Finally, the coding vectors then have one coefficient for every packet, not every symbol. Note that the same protocol and analysis of \cite{infocom09} holds even if we fix the basic unit of data as a symbol instead of a packet. The problem is that the complexity will be very high as the size of the coding matrix will be related to the number of symbols in the coding buffer, which is much more than the number of packets (typically, a symbol is one byte long).

In actual practice, TCP is a byte-stream oriented protocol in which ACKs are in terms of byte sequence numbers. If all packets are of fixed length, we can still apply the packet-level approach, since we have a clear and consistent map between packet sequence numbers and byte sequence numbers. In reality, however, TCP might generate segments of different sizes. The choice of how many bytes to group into a segment is usually made based on the Maximum Transmission Unit (MTU) of the network, which could vary with time. A more common occurrence is that applications may use the PUSH flag option asking TCP to packetize the currently outstanding bytes into a segment, even if it does not form a segment of the maximum allowed size. In short, it is important to ensure that our protocol works correctly in spite of variable packet sizes.

A closely related problem is that of repacketization.  Repacketization, as described in Chapter 21 of \cite{TCPbook1}, refers to the situation where a set of bytes that were assigned to two different segments earlier by TCP may later be reassigned to the same segment during retransmission.  As a result, the grouping of bytes into packets under TCP may not be fixed over time.

Both variable packet lengths and repacketization need to be addressed when implementing the coding protocol. To solve the first problem, if we have packets of different lengths, we could elongate the shorter packets by appending sufficiently many dummy zero symbols until all packets have the same length. This will work correctly as long as the receiver is somehow informed how many zeros were appended to each packet. While transmitting these extra dummy symbols will decrease the throughput, generally this loss will not be significant, as packet lengths are usually consistent.  

However, if we have repacketization, then we have another problem, namely it is no longer possible to view a packet as a single unknown. This is because we would not have a one-to-one mapping between packets sequence numbers and byte sequence numbers;  the same bytes may now occur in more than one packet. Repacketization appears to destroy the convenience of performing coding and decoding at the packet level.

To counter these problems, we propose the following solution. The coding operation described in \cite{infocom09} involves the sender storing the packets generated by the TCP source in a \emph{coding buffer}. We pre-process any incoming TCP segment before adding it to the coding buffer as follows:
\begin{enumerate}
	\item First, any part of the incoming segment that is already in the buffer is removed from the segment. 
	\item Next, a separate TCP packet is created out of each remaining contiguous part of the segment.
	\item The source and destination port information is removed. It will be added later in the network coding header. 
	\item The packets are appended with sufficiently many dummy zero bytes, to make them as long as the longest packet currently in the buffer. 
\end{enumerate}  
Every resulting packet is then added to the buffer. This processing ensures that the packets in the buffer will correspond to disjoint and contiguous sets of bytes from the byte stream, thereby restoring the one-to-one correspondence between the packet numbers and the byte sequence numbers. The reason the port information is excluded from the coding is because port information is necessary for the receiver to identify which TCP connection a coded packet corresponds to. Hence, the port information should not be involved in the coding. We refer to the remaining part of the header as the TCP subheader.

\begin{figure}%
\begin{center}
\includegraphics[width=\columnwidth]{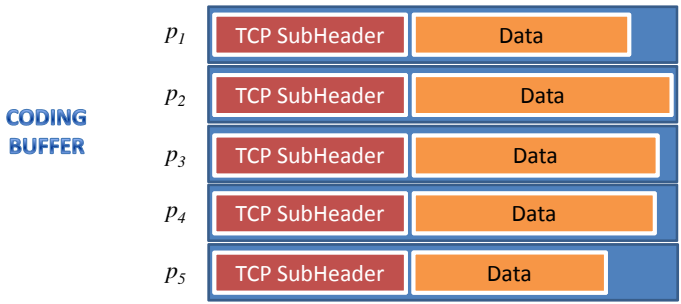}%
\caption{The coding buffer}%
\end{center}
\label{fig:codingbuffer}%
\end{figure}

Upon decoding the packet, the receiver can find out how many bytes are real and how many are dummy using the $Start_i$ and $End_i$ header fields in the network coding header (described below). With these fixes in place, we are ready to use the packet-level algorithm of \cite{infocom09}. All operations are performed on the packets in the coding buffer. Figure \ref{fig:codingbuffer} shows a typical state of the buffer after this pre-processing. The gaps at the end of the packets correspond to the appended zeros.
It is important to note that, as suggested in \cite{infocom09}, the TCP control packets such as SYN packet and reset packet are allowed to bypass the coding buffer and are directly delivered to the receiver without any coding.

\subsection{The coding header}
 A coded packet is created by forming a random linear combination of a subset of the packets in the coding buffer. The coding operations are done over a field of size 256 in our implementation. In this case, a field symbol corresdponds to one byte. The header of a coded packet should contain information that the receiver can use to identify what is the linear combination corresponding to the packet. We now discuss the header structure in more detail. 

We assume that the network coding header has the structure shown in Figure \ref{fig:codingheader}. The typical sizes (in bytes) of the various fields are written above them. The meaning of the various fields are described next:
\begin{figure*}%
\centering
\includegraphics[width=1.5\columnwidth]{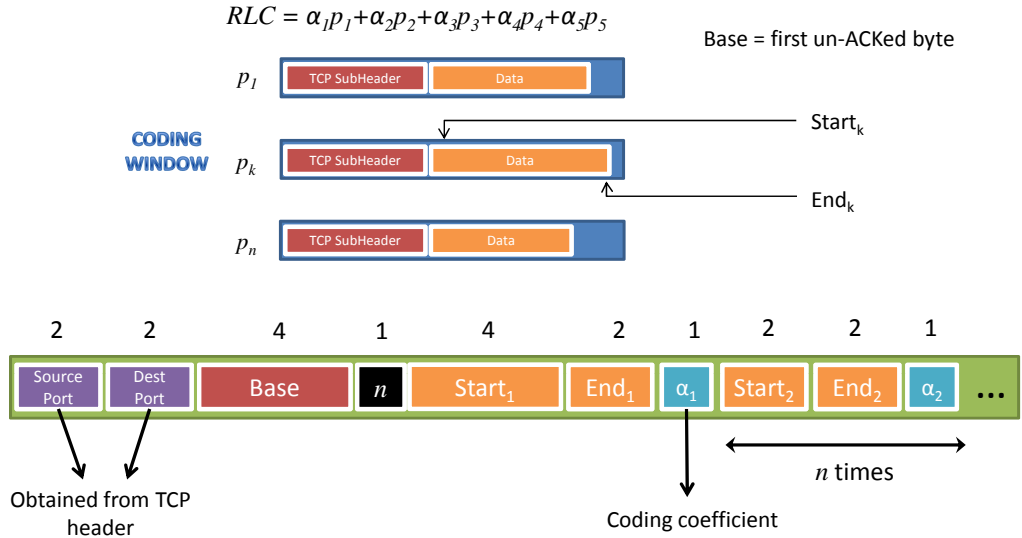}%
\caption{The network coding header}%
\label{fig:codingheader}%
\end{figure*}

\begin{itemize}
\item {\it Source and destination port:} The port information is needed for the receiver to identify the coded packet's session. It must not be included in the coding operation. It is taken out of the TCP header and included in the network coding header. 
\item {\it Base:} The TCP byte sequence number of the first byte that has not been ACKed. The field is used by intermediate nodes or the decoder to decide which packets can be safely dropped from their buffers without affecting reliability. 
\item {\it n:} The number of packets involved in the linear combination.
\item {\it $Start_i$:} The starting byte of the $i^{th}$ packet involved in the linear combination.
\item {\it $End_i$:} The last byte of the $i^{th}$ packet involved in the linear combination.
\item {\it $\alpha_i$:} The coefficient used for the $i^{th}$ packet involved in the linear combination.
\end{itemize}

The $Start_i$ (except $Start_1$) and $End_i$ are expressed relative to the previous packet's $End$ and $Start$ respectively, to save header space. As shown in the figure, this header format will add $5n+7$ bytes of overhead for the network coding header in addition to the TCP header, where $n$ is the number of packets involved in a linear combination. (Note that the port information is not counted in this overhead, since it has been removed from the TCP header.) We believe it is possible to reduce this overhead by further optimizing the header structure. 

\subsection{The coding window}
In the theoretical version of the algorithm, the sender transmits a random linear combination of all packets in the coding buffer. However, as noted above, the size of the header scales with the number of packets involved in the linear combination. Therefore, mixing all packets currently in the buffer will lead to a very large coding header. 

To solve this problem, we propose mixing only a constant-sized subset of the packets chosen from within the coding buffer. We call this subset the \emph{coding window}. The coding window evolves as follows. The algorithm uses a fixed parameter for the maximum coding window size $W$. The coding window contains the packet that arrived most recently from TCP (which could be a retransmission), and the $(W-1)$ packets before it in sequence number, if possible. However, if some of the $(W-1)$ preceding packets have already been dropped, then the window is allowed to extend beyond the most recently arrived packet until it includes $W$ packets. 

Note that this limit on the coding window implies that the code is now restricted in its power to correct erasures and to combat reordering-related issues. The choice of $W$ will thus play an important role in the performance of the scheme. The correct value for $W$ will depend on the length of burst errors that the channel is expected to produce. Other factors to be considered while choosing $W$ are discussed in Section \ref{sec:factors}. In our experiment, we fixed $W$ based on trial and error. 

\subsection{Buffer management}
A packet is removed from the coding buffer if a TCP ACK has arrived requesting a byte beyond the last byte of that packet. If a new TCP segment arrives when the coding buffer is full, then the segment with the newest set of bytes must be dropped. This may not always be the newly arrived segment, for instance, in the case of a TCP retransmission of a previously dropped segment. 

\section{Receiver side module}\label{sec:receiver}
The decoder module's operations are outlined below. The main data structure involved is the decoding matrix, which stores the coefficient vectors corresponding to the linear combinations currently in the decoding buffer. 

\subsection{Acknowledgment}
The receiver side module stores the incoming linear combination in the decoding buffer. Then it unwraps the coding header and appends the new coefficient vector to the decoding matrix. Gaussian elimination is performed and the packet is dropped if it is not innovative (i.e. if it is not linearly independent of previously received linear combinations). After Gaussian elimination, the oldest unseen packet is identified. Instead of acknowledging the packet number as in \cite{infocom09}, the decoder acknowledges the last seen packet by \emph{requesting the byte sequence number of the first byte of the first unseen packet}, using a regular TCP ACK. Note that this could happen before the packet is decoded and delivered to the receiver TCP. The port and IP address information for sending this ACK may be obtained from the SYN packet at the beginning of the connection. Any ACKs generated by the receiver TCP are not sent to the sender. They are instead used to update the receive window field that is used in the TCP ACKs generated by the decoder (see subsection below). They are also used to keep track of which bytes have been delivered, for buffer management. 

\subsection{Decoding and delivery}
The Gaussian elimination operations are performed not only on the decoding coefficient matrix, but correspondingly also on the coded packets themselves. When a new packet is decoded, any dummy zero symbols that were added by the encoder are pruned using the coding header information. A new TCP packet is created with the newly decoded data and the appropriate TCP header fields and this is then delivered to the receiver TCP. 

\subsection{Buffer management}
The decoding buffer needs to store packets that have not yet been decoded and delivered to the TCP receiver. Delivery can be confirmed using the receiver TCP's ACKs. In addition, the buffer also needs to store those packets that have been delivered but have not yet been dropped by the encoder from the coding buffer. This is because, such packets may still be involved in incoming linear combinations. The $Base$ field in the coding header addresses this issue. $Base$ is the oldest byte in the coding buffer. Therefore, the decoder can drop a packet if its last byte is smaller than $Base$, and in addition, has been delivered to and ACKed by the receiver TCP. Whenever a new linear combination arrives, the value of $Base$ is updated from the header, and any packets that can be dropped are dropped. 

\begin{figure}%
\centering
\includegraphics[width=\columnwidth]{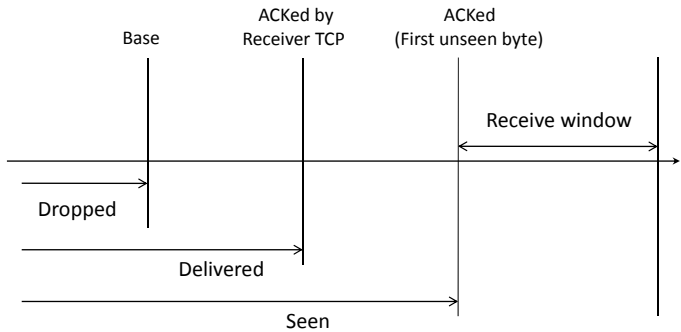}%
\caption{Receiver side window management}%
\label{fig:windows}%
\end{figure}

The buffer management can be understood using Fig. \ref{fig:windows}. It shows the receiver side windows in a typical situation. In this case, $Base$ is less than the last delivered byte. Hence, some delivered packets have not yet been dropped. There could also be a case where $Base$ is beyond the last delivered byte, possibly because nothing has been decoded in a while. 

\subsection{Modifying the receive window}
The TCP receive window header field is used by the receiver to inform the sender how many bytes it can accept. Since the receiver TCP's ACKs are suppressed, the decoder must copy this information in the ACKs that it sends to the sender. However, to ensure correctness, we may have to modify the value of the TCP receive window based on the decoding buffer size. The last acceptable byte should thus be the minimum of the receiver TCP's last acceptable byte and the last byte that the decoding buffer can accommodate. Note that while calculating the space left in the decoding buffer, we can include the space occupied by data that has already been delivered to the receiver because such data will get dropped when $Base$ is updated. If window scaling option is used by TCP, this needs to be noted from the SYN packet, so that the modified value of the receive window can be correctly reported. Ideally, we would like to choose a large enough decoding buffer size so that the decoding buffer would not be the bottleneck and this modification would never be needed. 

\section{Discussion of the practicalities}\label{sec:factors}
\subsection{Redundancy factor}
The choice of redundancy factor is based on the effective loss probability on the links. For a loss rate of $p_e$, with an infinite window W and using TCP Vegas, the theoretically optimal value of $R$ is $1/(1-p_e)$, as shown in \cite{infocom09}. The basic idea is that of the coded packets that are sent into the network, only a fraction $(1-p_e)$ of them are delivered on average. Hence, the value of $R$ must be chosen so that in spite of these losses, the receiver is able to collect linear equations at the same rate as the rate at which the unknown packets are mixed in them by the encoder. As discussed below, in practice, the value of $R$ may depend on the coding window size $W$. As $W$ decreases, the erasure correction capability of the code goes down. Hence, we may need a larger $R$ to compensate and ensure that the losses are still masked from TCP. Another factor that affects the choice of $R$ is the use of TCP Reno. The TCP Reno mechanism causes the transmission rate to fluctuate around the link capacity, and this leads to some additional losses over and above the link losses. Therefore, the optimal choice of $R$ may be higher than $1/(1-p_e)$. 

\subsection{Coding Window Size}
There are several considerations to keep in mind while choosing $W$, the coding window size The main idea behind coding is to mask the losses on the channel from TCP. In other words, we wish to correct losses without relying on the ACKs. Consider a case where $W$ is just 1. Then, this is a simple repetition code. Every packet is repeated $R$ times on average. Now, such a repetition would be useful only for recovering one packet, if it was lost. Instead, if $W$ was say 3, then every linear combination would be useful to recover any of the three packets involved. Ideally, the linear combinations generated should be able to correct the loss of any of the packets that have not yet been ACKed. For this, we need $W$ to be large. This may be difficult, since a large $W$ would lead to a large coding header. Another penalty of choosing a large value of $W$ is related to the interaction with TCP Reno. This is discussed in the next subsection. 

The penalty of keeping $W$ small on the other hand, is that it reduces the error correction capability of the code. For a loss probability of 10\%, the theoretical value of  $R$ is around 1.1. However, this assumes that all linear combinations are useful to correct any packet's loss. The restriction on $W$ means that a coded packet can be used only for recovering those $W$ packets that have been mixed to form that coded packet. In particular, if there is a contiguous burst of losses that result in a situation where the receiver has received no linear combination involving a particular original packet, then that packet will show up as a loss to TCP. This could happen even if the value of $R$ is chosen according to the theoretical value. To compensate, we may have to choose a larger $R$. 

The connection between $W$, $R$ and the losses that are visible to TCP can be visualized as follows. Imagine a process in which whenever the receiver receives an innovative linear combination, one imaginary token is generated, and whenever the sender slides the coding window forward by one packet, one token is used up. If the sender slides the coding window forward when there are no tokens left, then this leads to a packet loss that will be visible to TCP. The reason is, when this happens, the decoder will not be able to see the very next unseen packet in order. Instead, it will skip one packet in the sequence. This will make the decoder generate duplicate ACKs requesting that lost (i.e., unseen) packet, thereby causing the sender to notice the loss.

In this process, $W$ corresponds to the initial number of tokens available at the sender. Thus, when the difference between the number of redundant packets (linear equations) received and the number of original packets (unknowns) involved in the coding up to that point is less than $W$, the losses will be masked from TCP. However, if this difference exceeds $W$, the losses will no longer be masked. A theoretically optimal value of $W$ is not known. However, we expect that the value should be a function of the loss probability of the link. For the experiment, we chose values of $W$ based on trial and error. 

\subsection{Working with TCP Reno}
By adding enough redundancy, the coding operation essentially converts the lossiness of the channel into an extension of the round-trip time (RTT). This is why \cite{infocom09} proposed the use of the idea with TCP Vegas, since TCP Vegas controls the congestion window in a smoother manner using RTT,  compared to the more abrupt loss-based variations of TCP Reno. However, the coding mechanism is also compatible with TCP Reno. The choice of $W$ plays an important role in ensuring this compatibility. The choice of $W$ controls the power of the underlying code, and hence determines when losses are visible to TCP. As explained above, losses will be masked from TCP as long as the number of received equations is no more than $W$ short of the number of unknowns involved in them. For compatibility with Reno, we need to make sure that whenever the sending rate exceeds the link capacity, the resulting queue drops are visible to TCP as losses. A very large value of $W$ is likely to mask even these congestion losses, thereby temporarily giving TCP a false estimate of capacity. This will eventually lead to a timeout, and will affect throughput. The value of $W$ should therefore be large enough to mask the link losses and small enough to allow TCP to see the queue drops due to congestion. 

\subsection{Computational overhead}
It is important to implement the encoding and decoding operations efficiently, since any time spent in these operations will affect the round-trip time perceived by TCP. The finite field operations over GF(256) have been optimized using the approach of \cite{book:gf256}, which proposes the use of logarithms to multiply elements. Over GF(256), each symbol is one byte long. Addition in GF(256) can be implemented easily as a bitwise XOR of the two bytes. 

The main computational overhead on the encoder side is the formation of the random linear combinations of the buffered packets. The management of the buffer also requires some computation, but this is small compared to the random linear coding, since the coding has to be done on every byte of the packets. Typically, packets have a length $L$ of around 1500 bytes. For every linear combination that is created, the coding operation involves $LW$ multiplications and $L(W-1)$ additions over $GF(256)$, where $W$ is the coding window size. Note that this has to be done $R$ times on average for every packet generated by TCP. Since the coded packets are newly created, allocating memory for them could also take time. 

On the decoder side, the main operation is the Gaussian elimination. Note that, to identify whether an incoming linear combination is innovative or not, we need to perform Gaussian elimination only on the decoding matrix, and not on the coded packet. If it is innovative, then we perform the row transformation operations of Gaussian elimination on the coded packet as well. This requires $O(LW)$ multiplications and additions to zero out the pivot columns in the newly added row. The complexity of the next step of zeroing out the newly formed pivot column in the existing rows of the decoding matrix varies depending on the current size and structure of the matrix. Upon decoding a new packet, it needs to be packaged as a TCP packet and delivered to the receiver. Since this requires allocating space for a new packet, this could also be expensive in terms of time.

As we will see in the next section, the benefits brought by the erasure correction begin to outweigh the overhead of the computation and coding header for loss rates of about 3\%. This could be improved further by more efficient implementation of the encoding and decoding operations.

\subsection{Miscellaneous considerations}\label{sec:misc}
The TCP/NC protocol requires no modification in the basic features of the TCP protocol on either the sender side or the receiver side. However, other special features of TCP that make use of the ACKs in ways other than to report the next required byte sequence number, will need to be handled carefully. For instance, implementing the timestamp option in the presence of network coding across packets may require some thought. With TCP/NC, the receiver may send an ACK for a packet even before it is decoded. Thus, the receiver may not have access to the timestamp of the packet when it sends the ACK. Similarly, the TCP checksum field has to be dealt with carefully. Since a TCP packet is ACKed even before it is decoded, its checksum cannot be tested before ACKing. One solution is to implement a separate checksum at the network coding layer to detect errors. In the same way, the various other TCP options that are available have to be implemented with care to ensure that they are not affected by the premature ACKs. 

\section{Results}\label{sec:results}
We test the protocol on a TCP flow running over a single-hop wireless link. The transmitter and receiver are Linux machines equipped with a wireless antenna. The experiment is performed over 802.11a with a bit-rate of 6 Mbps and a maximum of 5 link layer retransmission attempts. RTS-CTS is disabled.

Our implementation uses the Click modular router \cite{click}. In order to control the parameters of the setup, we use the predefined elements of Click. Since the two machines are physically close to each other, there are very few losses on the wireless link. Instead, we artificially induce packet losses using the $RandomSample$ element. Note that these packet losses are introduced before the wireless link. Hence, they will not be recovered by the link layer retransmissions, and have to be corrected by the layer above IP. The round-trip delay is empirically observed to be in the range of a few tens of milliseconds. The encoder and decoder queue sizes are set to 100 packets, and the size of the bottleneck queue just in front of the wireless link is set to 5 packets. In our setup, the loss inducing element is placed before the bottleneck queue. 

The quantity measured during the experiment is the goodput over a 20 second long TCP session. The goodput is measured using $iperf$ \cite{iperf}.  Each point in the plots shown is averaged over 4 or more iterations of such sessions, depending on the variability. Occasionally, when the iteration does not terminate and the connection times out, the corresponding iteration is neglected in the average, for both TCP and TCP/NC. This happens around 2 \% of the time, and is observed to be because of an unusually long burst of losses in the forward or return path. In the comparison, neither TCP nor TCP/NC uses selective ACKs. TCP uses delayed ACKs. However, we have not implemented delayed ACKs in TCP/NC at this point. 

\begin{figure}%
\centering
\includegraphics[width=\columnwidth]{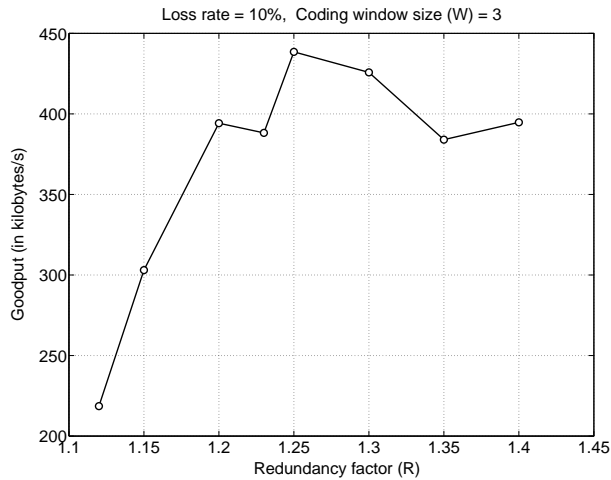}%
\caption{Goodput versus redundancy factor for a 10\% loss rate and W=3}%
\label{fig:redund}%
\end{figure}

Fig. \ref{fig:redund} shows the variation of the goodput with the redundancy factor $R$ for a loss rate of 10\%, with a fixed coding window size of $W=3$. The theoretically optimal value of $R$ for this loss rate is 1.11 (=1/0.9) (see \cite{infocom09}). However, from the experiment, we find that the best goodput is achieved for an $R$ of around 1.25. The discrepancy is possibly because of the type of coding scheme employed. Our coding scheme transmits a linear combination of only the $W$ most recent arrivals, in order to save packet header space. This restriction reduces the strength of the code for the same value of $R$. In general, the value of $R$ and $W$ must be carefully chosen to get the best benefit of the coding operation. As mentioned earlier, nother reason for the discrepancy could be the use of TCP Reno. 

\begin{figure}%
\centering
\includegraphics[width=\columnwidth]{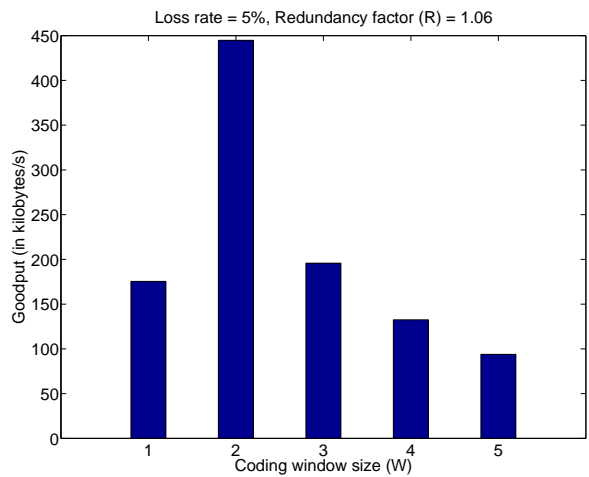}%
\caption{Goodput versus coding window size for a 5\% loss rate and R=1.06}%
\label{fig:codingwindow}%
\end{figure}

Fig. \ref{fig:codingwindow} plots the variation of goodput with the size of the coding window size $W$. The loss rate for this plot is 5\%, with the redundancy factor fixed at 1.06. We see that the best coding window size is 2. Note that a coding window size of $W=1$ corresponds to a repetition code that simply transmits every packet 1.06 times on average. In comparison, a simple sliding window code with $W=2$ brings a big gain in throughput by making the added redundancy more useful. However, going beyond 2 reduces the goodput because a large value of $W$ can mislead TCP into believing that the capacity is larger than it really is, which leads to timeouts. We find that the best value of $W$ for our setup is usually 2 for a loss rate up to around 5 \%, and is 3 for higher loss rates up to 25\%. Besides the loss rate, the value of $W$ could also depend on other factors such as the round-trip time of the path.   

\begin{figure}%
\centering
\includegraphics[width=\columnwidth]{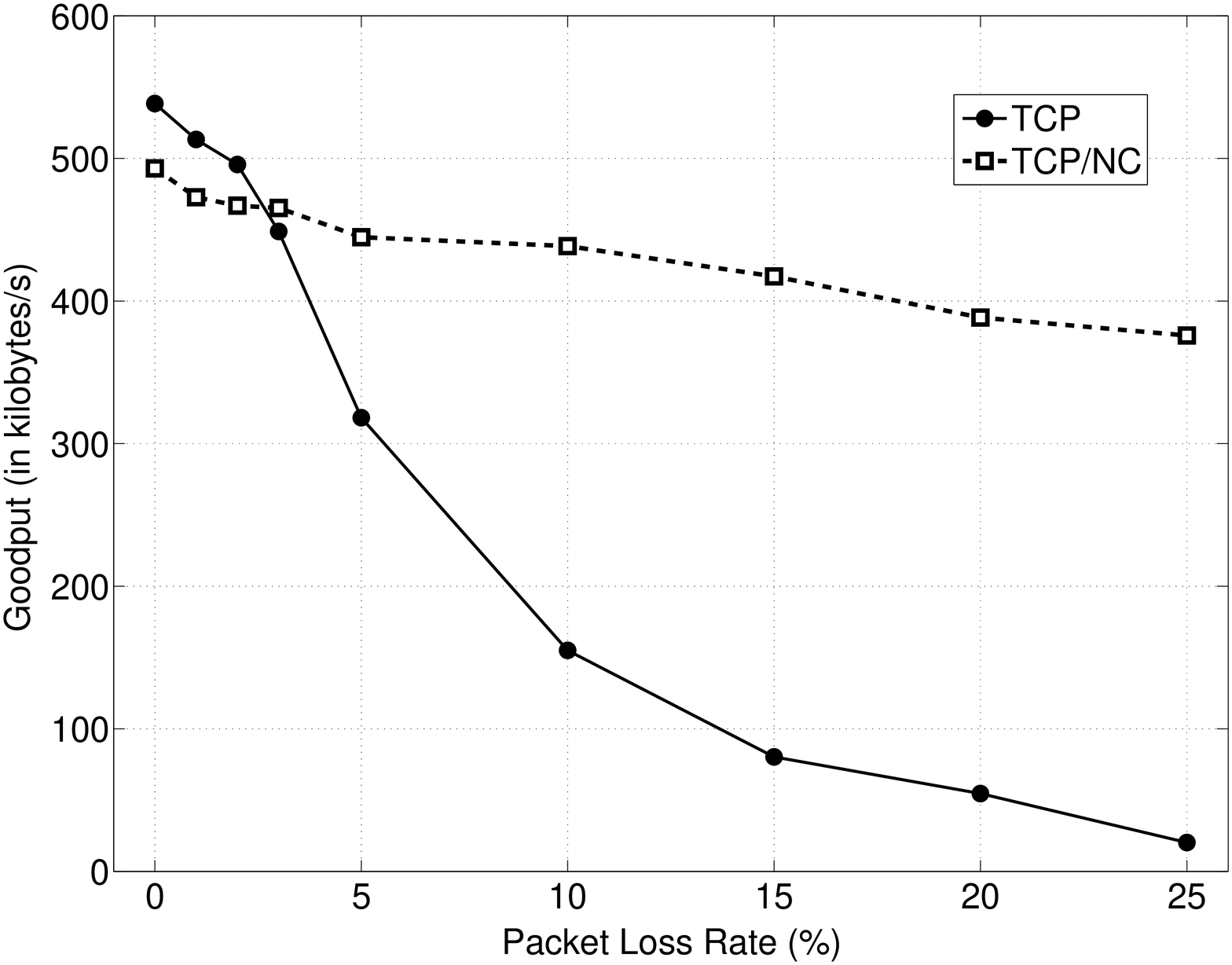}%
\caption{Goodput versus loss rate}%
\label{fig:lossrate}%
\end{figure}

Fig. \ref{fig:lossrate} shows the goodput as a function of the packet loss rate. For each loss rate, the values of $R$ and $W$ have been chosen by trial and error, to be the one that maximizes the goodput. We see that in the lossless case, TCP performs better than TCP/NC. This could be because of the computational overhead that is introduced by the coding and decoding operations, and also the coding header overhead. However, as the loss rate increases, the benefits of coding begin to outweigh the overhead. The goodput of TCP/NC is therefore higher than TCP. Coding allows losses to be masked from TCP, and hence the fall in goodput is more gradual with coding than without. The performance can be improved further by improving the efficiency of the computation.

\section{Conclusion and future work}\label{sec:conc}
In this paper, we have demonstrated experimentally, a new sliding-window coding scheme that is compatible with TCP-Reno. The scheme allows the interfacing of TCP with network coding. This has implications for running TCP over wireless networks, in particular in the context of lossy multipath opportunistic routing scenarios. We believe that the proposed ideas and the implementation will lead to the practical realization of the theoretically promised benefits of network coding in such scenarios.

This endeavor would require more work in the future, in terms of understanding the role played by the various parameters of the new protocol, for instance, the redundancy factor $R$ and the coding window size $W$. To achieve high throughputs in a fair manner, the values of $R$ and $W$ have to be carefully adapted based on the characteristics of the underlying channel. In the future, it would be useful to extend this implementation to a multi-hop network with multiple paths from the sender to the receiver and re-encoding of packets inside the network.

\section*{Acknowledgments}
The authors would like to thank Prof. Devavrat Shah and Prof. Dina Katabi for several useful discussions. We would also like to thank Mythili Vutukuru and Rahul Hariharan for their help and advice regarding the implementation.

\bibliography{refs}

\newcommand{\noopsort}[1]{} \newcommand{\printfirst}[2]{#1}
  \newcommand{\singleletter}[1]{#1} \newcommand{\switchargs}[2]{#2#1}
\begin{thebibliography}{10}
\providecommand{\url}[1]{#1}
\csname url@rmstyle\endcsname
\providecommand{\newblock}{\relax}
\providecommand{\bibinfo}[2]{#2}
\providecommand\BIBentrySTDinterwordspacing{\spaceskip=0pt\relax}
\providecommand\BIBentryALTinterwordstretchfactor{4}
\providecommand\BIBentryALTinterwordspacing{\spaceskip=\fontdimen2\font plus
\BIBentryALTinterwordstretchfactor\fontdimen3\font minus
  \fontdimen4\font\relax}
\providecommand\BIBforeignlanguage[2]{{%
\expandafter\ifx\csname l@#1\endcsname\relax
\typeout{** WARNING: IEEEtran.bst: No hyphenation pattern has been}%
\typeout{** loaded for the language `#1'. Using the pattern for}%
\typeout{** the default language instead.}%
\else
\language=\csname l@#1\endcsname
\fi
#2}}

\bibitem{rangwala}
S.~Rangwala, A.~Jindal, K.-Y. Jang, K.~Psounis, and R.~Govindan,
  ``Understanding congestion control in multi-hop wireless mesh networks,'' in
  \emph{Proc. of ACM/IEEE International Conference on Mobile Computing and
  Networking (MobiCom)}, 2008.

\bibitem{infocom09}
J.~K. Sundararajan, D.~Shah, M.~M\'edard, M.~Mitzenmacher, and J.~Barros,
  ``Network coding meets {TCP},'' in \emph{Proceedings of IEEE INFOCOM}, April
  2009, pp. 280--288.

\bibitem{Brockners99}
F.~Brockners, ``The case for {FEC}-fueled {TCP}-like congestion control,'' in
  \emph{Kommunikation in Verteilten Systemen}, ser. Informatik Aktuell,
  R.~Steinmetz, Ed.\hskip 1em plus 0.5em minus 0.4em\relax Springer, 1999, pp.
  250--263.

\bibitem{exor}
S.~Biswas and R.~Morris, ``{ExOR}: opportunistic multi-hop routing for wireless
  networks,'' in \emph{Proceedings of ACM SIGCOMM 2005}.\hskip 1em plus 0.5em
  minus 0.4em\relax ACM, 2005, pp. 133--144.

\bibitem{more}
S.~Chachulski, M.~Jennings, S.~Katti, and D.~Katabi, ``Trading structure for
  randomness in wireless opportunistic routing,'' in \emph{Proc. of ACM SIGCOMM
  2007}, August 2007.

\bibitem{RLC}
T.~Ho, M.~M\'edard, R.~Koetter, D.~Karger, M.~Effros, J.~Shi, and B.~Leong, ``A
  random linear network coding approach to multicast,'' \emph{IEEE Trans. on
  Information Theory}, vol.~52, no.~10, pp. 4413--4430, October 2006.

\bibitem{TCPbook1}
W.~R. Stevens, \emph{{TCP/IP} Illustrated, Volume 1: The Protocols}.\hskip 1em
  plus 0.5em minus 0.4em\relax Addison-Wesley, 1994.

\bibitem{book:gf256}
\BIBentryALTinterwordspacing
N.~R. Wagner, \emph{The Laws of Cryptography with Java Code}. [Online].
  Available: \url{{http://www.cs.utsa.edu/~wagner/lawsbookcolor/laws.pdf}}
\BIBentrySTDinterwordspacing

\bibitem{click}
E.~Kohler, R.~Morris, B.~Chen, J.~Jannotti, and M.~F. Kaashoek, ``The {C}lick
  modular router,'' in \emph{ACM Transactions on Computer Systems}, vol.~18,
  no.~3, August 2000, pp. 263--297.

\bibitem{iperf}
\BIBentryALTinterwordspacing
{NLANR Distributed Applications Support Team}, ``Iperf –- the tcp/udp bandwidth
  measurement tool.'' [Online]. Available:
  \url{{http://dast.nlanr.net/Projects/Iperf/}}
\BIBentrySTDinterwordspacing

\end{thebibliography}
\bibliographystyle{IEEEtran}
\end{document}